\font\sc=cmcsc10

\def\kpc{\;{\rm kpc} }

\def\Msun{\,M_\odot}
\def\Lsun{{\rm L}_{\rm \odot}}
\def\MLsun{\Msun/\Lsun}
\def\kms{\;{\rm km}\,{\rm s}^{-1}  }

\def\vdisp{\left<v^2\right>^{1/2}}
\def\vdispsqr{\left<v^2\right>}

\def\spose#1{\hbox to 0pt{#1\hss}}
\def\lta{\mathrel{\spose{\lower 3pt\hbox{$\sim$}}
    \raise 2.0pt\hbox{$<$}}}
\def\gta{\mathrel{\spose{\lower 3pt\hbox{$\sim$}}
    \raise 2.0pt\hbox{$>$}}}
\documentclass{aastex}
\usepackage{emulateapj5}

\begin{document}

\title{Ursa Major: a missing low-mass CDM halo?}
\author{Jan T. Kleyna$^1$, Mark I. Wilkinson$^2$, N. Wyn Evans$^2$, 
Gerard Gilmore$^2$}
\affil{$^1$Institute for Astronomy, 2680 Woodlawn Drive, Honolulu, HI, 96822, USA}
\affil{$^2$Institute of Astronomy, Madingley Road, Cambridge, CB3 OHA, UK}

\begin{abstract} 
The recently discovered Ursa Major dwarf spheroidal (dSph) galaxy
candidate is about five to eight times less luminous than the faintest
previously known dSphs And IX, Draco, and UMi.  In this Letter, we
present velocity measurements of seven color-magnitude selected UMajor
candidate stars. Two of them are apparent non-members based on
metallicity and velocity, and the remaining five stars yield a
systemic heliocentric velocity of $\bar v=-52.45\pm4.27\,\kms$ and a
central line of sight velocity dispersion of
$\vdisp=9.3^{+11.7}_{-1.2}\,\kms$, with 95\% confidence that
$\vdisp>6.5\,\kms$.  Assuming that UMajor is in dynamical equilibrium,
it is clearly dark matter dominated, and cannot be a purely stellar
system like a globular cluster.  It has an inferred central
mass-to-light ratio of $M/L\sim 500\,\MLsun$ and, based on our
studies of other dSphs, may possess a much larger total mass to light
ratio. UMajor is unexpectedly massive for its low luminosity --
indeed, UMajor appears to be the most dark-matter dominated galaxy yet
discovered. The presence of so much dark matter in UMajor immediately
suggests that it may be a member of the missing population of low-mass
galaxies predicted by the Cold Dark Matter (CDM) paradigm.  Given the
weak correlation between dSph mass and luminosity, it is entirely
likely that a population of dark dwarfs surrounds our Galaxy.
\end{abstract}

\keywords{galaxies: individual: Ursa Major dwarf -- galaxies:
kinematics and dynamics -- Local Group -- dark matter -- celestial
mechanics, stellar dynamics }

\section{INTRODUCTION}

All of the Local Group dwarf spheroidal (dSph) galaxies have velocity
dispersions much larger than expected for self-gravitating stellar
systems, implying that their dynamics is dominated by dark matter,
with the stars being little more than dynamical tracers within a dark
halo \citep[{\it e.g.} ][]{MAT98}.  Some dSphs have central mass to
light ratios of $\sim 100\,\MLsun$, and average mass to light ratios
of several hundred
\citep[{\it e.g.} ][]{KleynaDraco, KleynaSextans, WilkDracoUMi}.

Recently, \citet{Willman05} discovered a new candidate dSph in Ursa
Major in a search of data from the Sloan Digital Sky Survey
(SDSS) \citep{ABA05}. Located about $100\kpc$ from the Galaxy, its
half-light radius is $r_{1/2}\sim250\,\rm pc$, covering $7.75^\prime$
on the sky.  Most remarkably, with $M_V\sim-6.75$, it is about five
times less luminous than the faintest previously known dSph
And\,IX \citep{Zucker04, Chapman05} and about eight times less
luminous than the faintest Galactic dSphs UMi and Draco \citep{MAT98}.


In this paper, we describe precise radial velocity measurements of
seven UMajor candidate stars.  We compute UMajor's inferred central
density and mass, and discuss how UMajor fits into the rest of the
rest of the dSph population within the context of standard CDM theory.

\section{DATA}

\begin{figure}
\begin{center}
\plotone{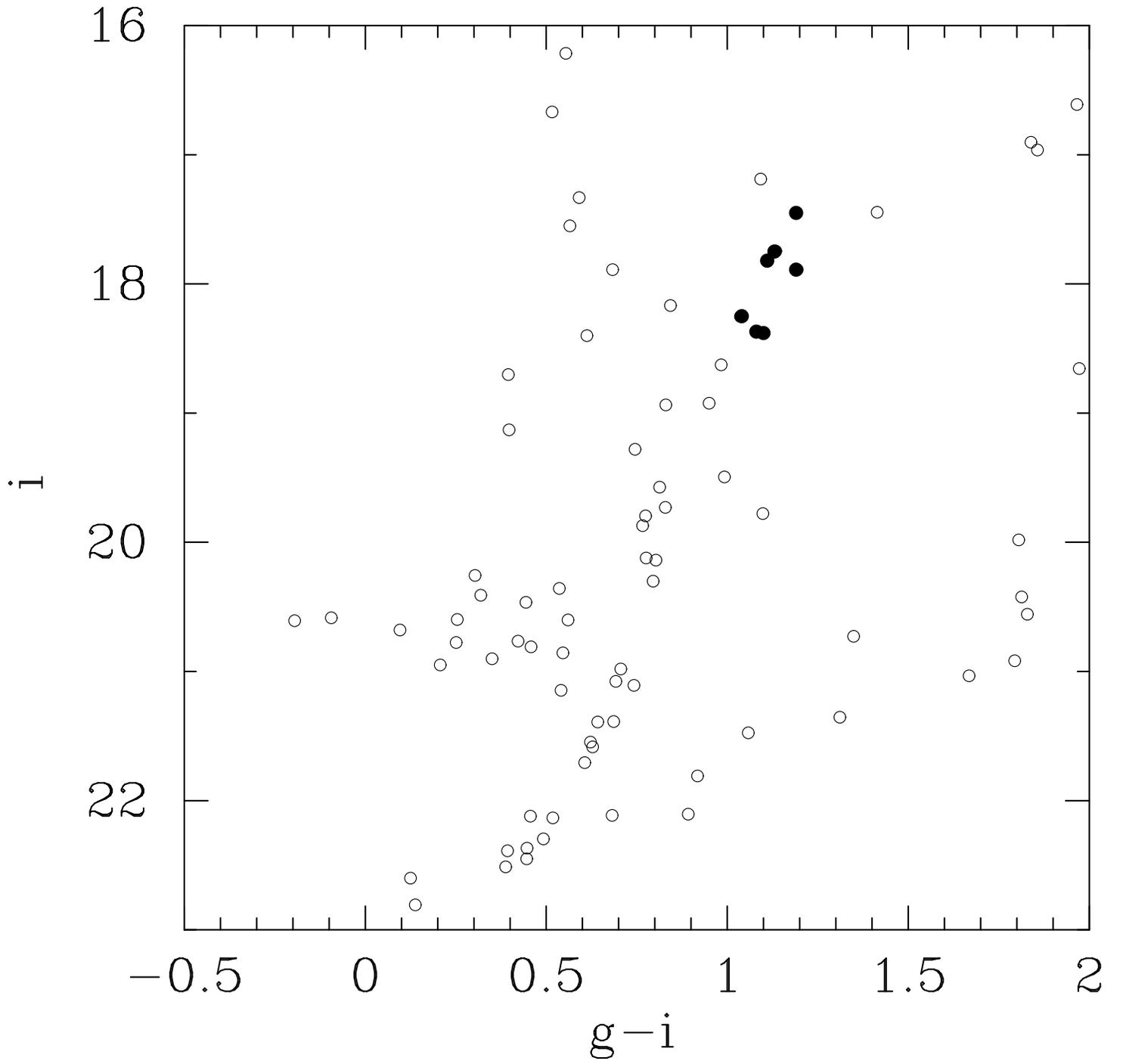}        
\end{center}
\caption{\label{fig:cmd}SDSS color-magnitude diagram for stars in the inner $6^\prime$ of 
Ursa Major, with clearly visible giant and horizontal branches. The 
filled circles represent giant branch candidates observed in this work.}
\end{figure}

We selected candidate stars from Data Release Three of SDSS, picking
targets within $6^\prime$ of UMajor's position as given
by \citet{Willman05}.  Figure \ref{fig:cmd} shows the color magnitude
diagram of the central $6^\prime$ region, slightly smaller than
Willman's half-light radius $r_{1/2}=7.75^\prime$. The giant branch
and horizontal branch are both visible, although together they contain
only about 50 stars.  Our seven targets (dark circles in
Figure \ref{fig:cmd}) were drawn from the brightest part of the giant
branch, and span the magnitude range $i=17.45$ to $i=18.38$.
Additionally, we observed the bright velocity standards HD107328,
HD90861, and HD132737.

We observed our stars using the upgraded HIRES \citep{Vogt94} echelle
spectrograph on the Keck I telescope on the night of May 17,
2005. Each star was observed in one integration lasting 1800s.  The
long spectral coverage of the spectrograph allowed us to obtain
wavelengths from H$\alpha$ ($\rm 6564\AA$) to the red-most line of the
Calcium triplet ($\rm 8662\AA$).  The signal to noise varied among the
spectra from $S/N=12$ to 3 per pixel, or $S/N=61$ to 17 per \AA.  Only
the red-most Ca triplet line was near a sky line, but sky subtraction
was generally clean even in this case.

We extracted the spectra using the {\sc makee} data reduction package
for HIRES, creating flux and variance spectra for each echelle order.
Because the latest version of {\sc makee} for the new three-chip
upgrade to HIRES does not at present solve for the dispersion, we fit
the dispersion solution manually for the relevant echelle orders using
{\sc IRAF}.  All calibration arc exposures had to be taken during the
afternoon or morning because of persistent ghosting with the new
chips. To verify instrument stability, we measured the positions of
sky lines in the science exposures, and found them to be constant
within $0.2\,\kms$.

For each of the four stellar absorption lines of interest (H$\alpha$
and the three Ca lines), we cross-correlated a synthetic
Gaussian template with the appropriate echelle order using the {\sc
IRAF} {\sc fxcor} package.  In all cases except for the third Ca
triplet line for the faintest UMajor star, we obtained a clear
cross-correlation function peak.

A potential problem with single-slit observations is that
mis-centering of the star on the slit will produce a velocity offset
common to all orders.  To address this problem, we note that telluric
absorption lines experience the same spurious velocity shift as
stellar absorption lines, and can be used to adjust the velocity back
to its correct value. Accordingly, we computed a velocity adjustment
using telluric absorption features around $\rm 6880\AA$.  Using a
template created from the telluric lines from one of the bright
standard stars, we used cross-correlation to compute relative velocity
adjustments of the other stars.  In all cases, the adjustment was
less than $0.8\,\kms$.

An additional potential source of inaccuracy is template mismatch; for
instance, the H$\alpha$ line is actually a blend of several different
transitions, and may not be exactly at the template wavelength.  To
address this problem, we adjusted all of the velocities for each
stellar line en masse relative to the second Ca line by computing the
median velocity difference between the line and the second Ca line.
This procedure introduced a shift of at most $1.6\,\kms$.

Next, we treated each velocity for each stellar line as an independent
measurement, and combined them to obtain a final velocity and error.
Finally, we shifted the entire velocity set to bring the standards
into agreement with their published velocities, with a final scatter
of $0.4\,\kms$, slightly larger then the intrinsic standard star
uncertainty of $0.3\,\kms$. For each star, we obtained two error
estimates: $\sigma_{\rm IRAF}$, obtained by rescaling the nominal {\sc
IRAF} {\sc fxcor} velocity errors to give the expected $\chi^2$; and
$\sigma_{\rm scat}$, obtained from the empirical velocity scatter
among individual lines.  For UMajor stars, the two errors are
generally similar, but $\sigma_{\rm IRAF}$ apparently overestimates
the errors for the bright standard stars.  Table \ref{table:stars}
shows the results for our seven target stars.

\begin{figure}
\begin{center} 
\plotone{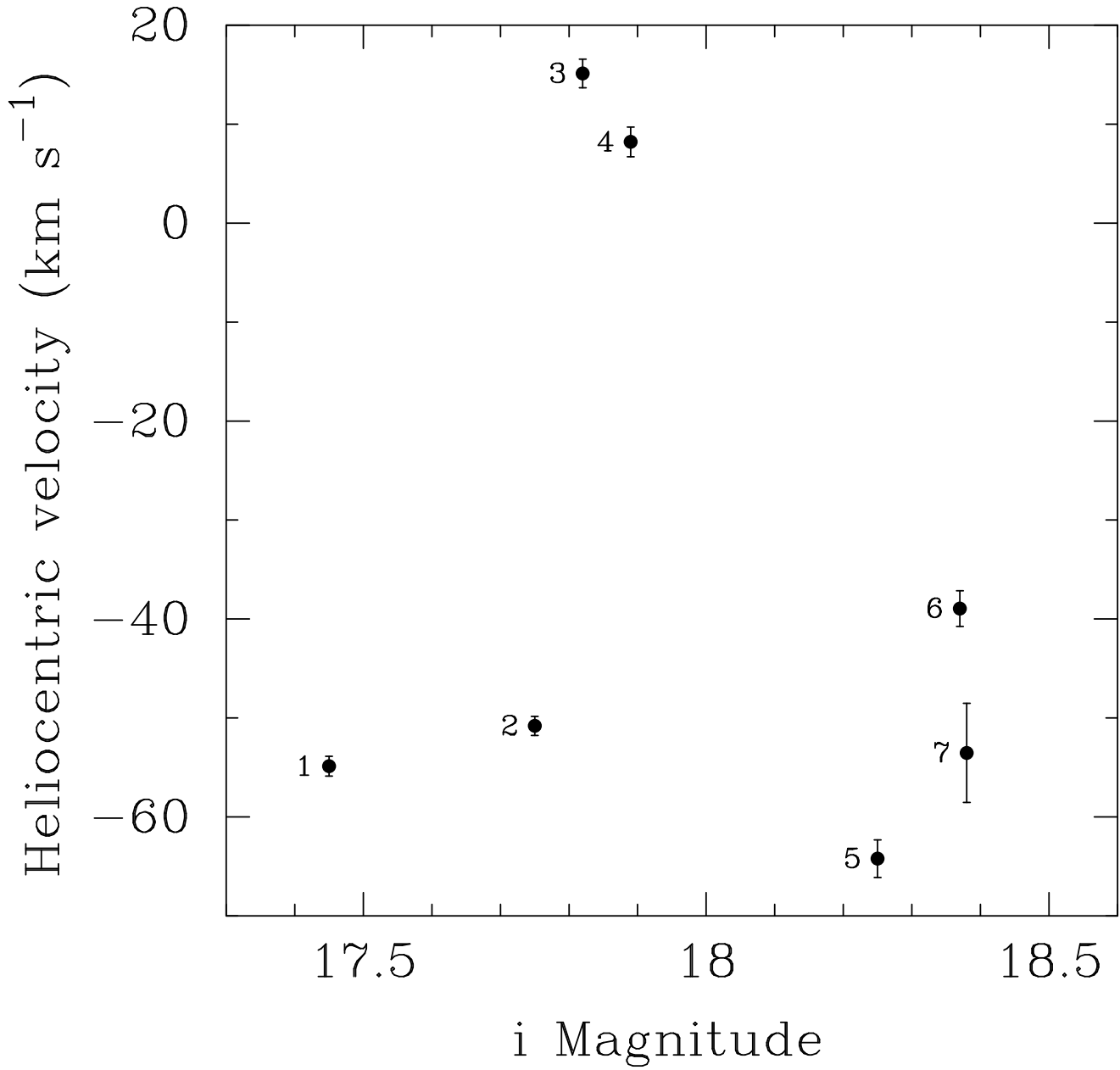}
\end{center}
\caption{\label{fig:vels}Velocities of Ursa Major candidate objects and
  $\sigma_{\rm scat}$ uncertainties. Stars 3 and 4 (near top) appear
to be non-members, with velocities displaced about $60\,\kms$ from the
others.}
\end{figure}

In Figure \ref{fig:vels}, we show the velocities as a function
of magnitude.  Objects 3 and 4 are clearly outliers, separated
about $60\,\kms$ from the other five objects. 

The equivalent width of the Ca triplet can be used as a measure of a
star's metallicity, for known surface gravity
(dwarf vs. giant) \citep[{\it e.g.} ][]{ARM91}. To establish whether the
kinematical outliers are members of the UMajor population, we computed
the equivalent width of the Ca triplet lines by fitting them with a
Lorentzian profile plus a linear continuum, and then integrating the
difference between the line profile fit and the continuum fit. We
compute the uncertainties of the equivalent widths using a Monte-Carlo
procedure: we generated simulated data using our best fit solution
added to the empirical noise spectrum output by the {\sc makee}
reduction package, and fit the simulated data in the same manner as
the real data.

\begin{figure}
\begin{center} 
\plotone{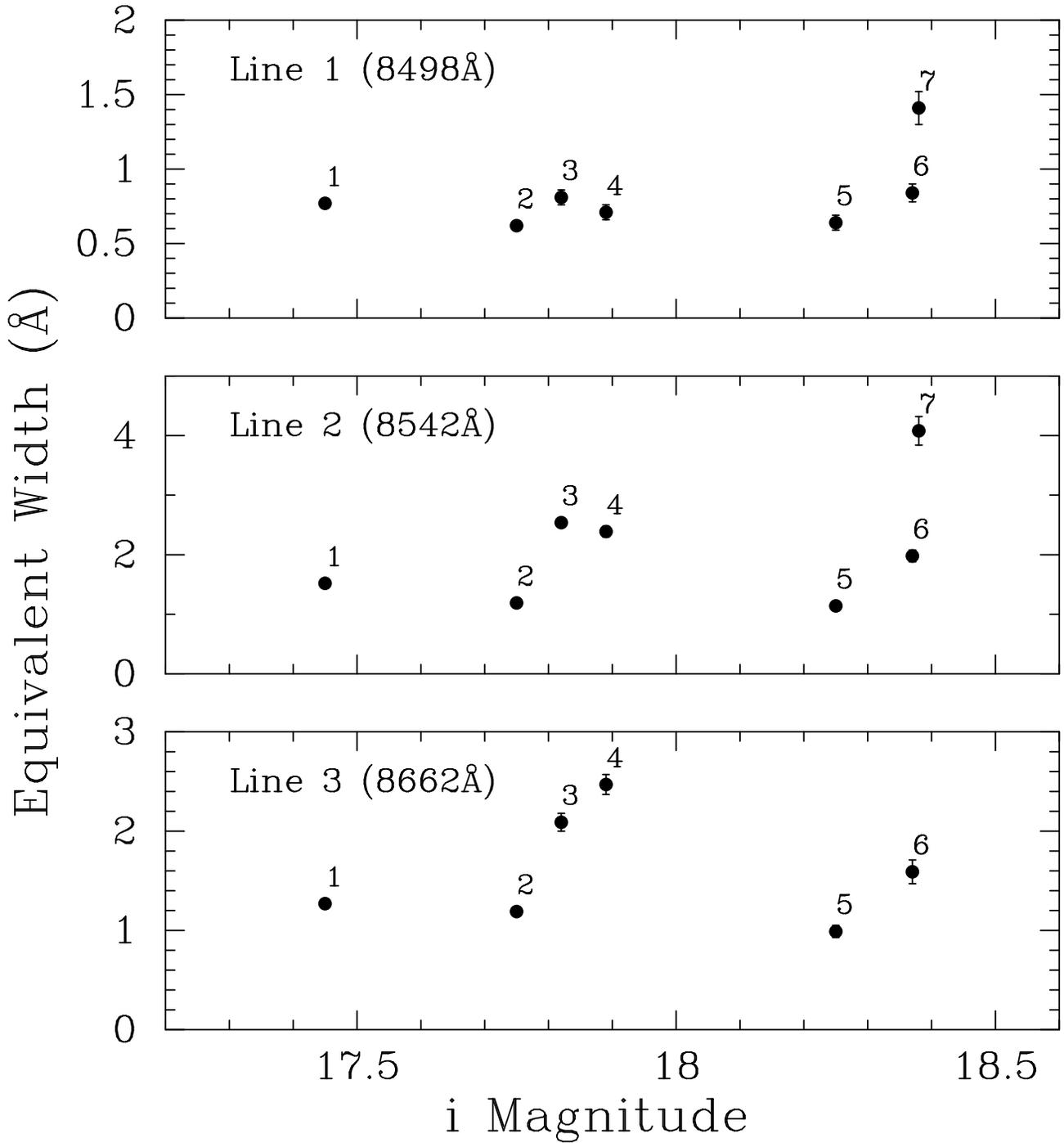}
\end{center}
\caption{\label{fig:ew}Equivalent widths of the three Calcium triplet lines. The reddest two lines of objects 3 and 4, which are apparent non-members on the basis of velocity, also have a significantly larger equivalent width than the other stars.  The equivalent widths for object 7 are more uncertain than the error bars indicate because the absorption lines were almost invisible to the eye, and the fits may be entirely spurious. No fit could be obtained for the reddest line of object 7.}
\end{figure}

Figure \ref{fig:ew} shows the equivalent widths for the three Ca
lines.  The kinematical outliers 3 and 4 clearly have a larger
equivalent width for the second and third Ca lines, although the first
line shows no difference. Object 7 produces very poor fits because the
lines are barely discernible above the noise, and its equivalent
widths are highly suspect.  Generally, only the second and third lines
are used for metallicity determination \citep{ARM91}, so the absence
of an effect for the weaker first line is not a source of concern.
However, for a fixed metallicity, the equivalent width is
anti-correlated with magnitude with a slope about $\rm 0.6\AA$ per
magnitude \citep{ARM91}, so that it appears anomalous that star 6 has
a larger equivalent width than the brighter stars 1, 2, and 5. To be
cautious, we consider the possibility that both 6 and 7 are
non-members.

\section{VELOCITY DISPERSION AND SYSTEMIC VELOCITY}

The simple RMS line of sight dispersion of the five member objects
(1,2,5,6, and 7) is $9.1\,\kms$, to which measurement errors
contribute only slightly.  To compute the dispersion $\vdisp$ more
rigorously, we assume that the objects are drawn from a Gaussian
distribution, and apply equation 2 of \citet{KleynaSextans}, imposing
an {\it a priori} maximum value of $50\,\kms$ on the dispersion, and
using $\sigma_{\rm scat}$ as the velocity uncertainty. We do not
consider the effect of binary stars, which is generally small.
Including all member stars, we obtain
$\vdisp=9.3^{+11.7}_{-1.2}\,\kms$, where the uncertainties are
$1\sigma$, defined such that 16\% of the probability distribution is
both above and below the error interval.  If we omit object 7, then we
have $\vdisp=10.4^{+16.9}_{-1.2}\,\kms$, and if we omit both 6 and 7,
$\vdisp=6.9^{+19.7}_{-0.4}\,\kms$.  For these three cases, in the same
order, we are confident with 95\% certainty that the dispersion is
greater than 6.5, 7.3, and 4.7 $\kms$.

Assuming five genuine members with a true dispersion of $9.3\,\kms$,
the systemic heliocentric velocity of UMajor is $-52.5\pm4.3 \, \kms$.

\section{MASS TO LIGHT RATIO}

The central mass to light ratio of a stellar system with an isotropic
orbital distribution and constant $M/L$ may be expressed as
$(M/L)_0=9\,\eta \left<v^2\right>/(2\pi G r_{hb} S_0)$, where
$\eta$ is a luminosity distribution dependent factor always close to
1, $r_{hb}$ is the half-light radius, and $S_0$ is the surface
brightness \citep{RT86}. Using UMajor's approximate published
total luminosity $M_V=-6.75$ ($L_V\sim4\times10^4\Lsun$) and
half-light radius $r_{1/2}=250\,\rm pc$ \citep{Willman05}, we obtain a
central surface brightness of $S_0=0.11\,\Lsun\rm pc^{-2}$.  Assuming
that $\vdisp=9.3$, one then obtains $(M/L)_0\approx 550 \MLsun$,
as one might expect for a system that is nearly an order of magnitude
fainter than known dark-matter dominated dSphs, but has the same
dispersion.  The central density is then $\rho_0=0.18\,\Msun\,\rm
pc^{-3}=6.8\,{\rm GeV\,cm^{-3}}$ using the above equation, or
$\rho_0=0.22\,\Msun\,\rm pc^{-3}$ using the formula
$\rho_0=166\vdispsqr/r_{\rm core}$ \citep{MAT98}.  Despite UMajor's
extreme $M/L$, its density is within a factor of two of that of
several other dSphs \citep{MAT98}. Several important caveats apply,
however. First, we have identified the half-light radius with the
half-brightness radius, when in truth they differ. If UMajor has a
Plummer profile, for example, then $r_{\rm hb}=0.64 \, r_{1/2}$, and
we underestimate $M/L$ substantially.  Next, the above formula for
$\rho_0/I_0$ strictly applies only to systems with constant $M/L$,
whereas a dwarf with a halo has varying $M/L$, a fact that is often
ignored when computing $M/L$ using core fitting.  Thus the above
values of $(M/L)_0$ and $\rho_0$ are intrinsically imprecise, and
are useful primarily as relative values to compare with other dSphs
having similarly computed qualities.

An alternative approach to computing the mass of the system is the
projected mass estimator \citep{Bahcall81, Heisler85, Evans03},
whereby the mass of a system with measured radial velocities $v_i$ at
projected radii $R_i$ is given by $M=C/(GN) \times \sum_{i=1}^N v_i^2
R_i$, where $C$ is a constant depending on the mass and light
distributions of the system.
For our five-member UMajor data, this becomes $M=C\,1.5\times10^6\Msun$.
\citet{Evans03} compute $C$ for the general case of measuring the mass
between radii $r_{\rm min}$ to $r_{\rm max}$ for a $\rho\propto
r^{-\gamma}$ tracer population in a $\phi\propto r^{-\alpha}$
potential.  Taking $r_{\rm min}=100\,\rm pc$ from our innermost star
and assuming that all the stars are within the half-light radius so
that $r_{\rm max}=250\,\rm pc$, and assuming that $\gamma=3$, with a
halo having a flat rotation curve ($\alpha=0$), we obtain $C\approx6$,
again giving $M\approx10^7\Msun$ and $M/L\sim 500$ inside the region
observed.  The mass varies from $5\times10^6\Msun$ to
$2\times10^7\Msun$ as $\gamma$ ranges from 2 to 6.

We further note that dSphs for which radially extended data exists
have a mean $M/L$ within the apparent stellar cutoff or ``tidal''
radius that is nearly an order of magnitude larger than the core-fit
central $M/L$ ({\it e.g.} compare \citet{MAT98}
with \citet{KleynaDraco},
\citet{WilkDracoUMi}, and \citet{KleynaSextans}).  If we accept the 
CDM result that halos are similar in structure, then UMajor might have
a global $M/L\sim 3000\MLsun$.

\section{DISCUSSION}

Our stellar velocity measurements show that UMajor is a nearly dark,
low-mass halo, almost an order of magnitude less luminous than
previously known dSph galaxies of similar mass.   We obtain a central
mass-to-light ratio $M/L\sim500\MLsun$; if this extrapolates to larger
radii like other dSphs, UMajor may have a mean $M/L\sim3000\MLsun$.
Hence, UMajor may be the most dark matter dominated galaxy yet
discovered. For comparison, the recently discovered $10^{11}\Msun$
VIRGOHI 21 H{\sc i} source has $M/L>500\MLsun$ \citep{Minchin05}, and
the extreme dark spiral NGC\,2915 has $M/L\approx80$ \citep{Meurer96}.

The CDM structure formation paradigm predicts that dark
matter clumps into cusped halos characterized by a single parameter,
the central density, with the smallest and least massive halos being
the densest \citep{NFW97}.  Problematically, CDM also predicts an order
of magnitude more small Galactic satellite halos than are actually
observed \citep[{\it e.g.} ][]{Klypin99}, but it is unclear whether
the theory is at fault, or whether the missing halos are too dark to 
be observed.



High Velocity Clouds (HVCs) have been suggested as candidates for the
missing CDM dwarf halos \citep{Blitz99}. This suggestion remains
controversial, especially as the high velocity clouds may themselves
be divided into compact and diffuse subclasses with possibly different
origins \citep[{\it e.g.} ][]{deHeij02}. There is no clear distance
determinant for the HVCs around the Milky Way, and, for a long time,
there was no clear consensus as to whether the HVCs were Galactic or
extra-Galactic. This ambiguity may have been resolved by the detection
of a faint circumgalactic HVC population around the Andromeda galaxy
by \citet{Thilker04}, who argue that the available data are consistent
with formation mechanisms via cooling flows, or with tidal debris from
mergers, or with gaseous counterparts of the missing CDM haloes.

Accordingly, UMajor may represent the best candidate for a ``missing''
CDM halo.  Its existence raises several interesting questions.

Why do the dSphs have a similar velocity dispersion, a similar central
density and, presumably, a similar mass, as noted
by \citet{MateoLeo98}?  It is curious that Draco, UMi, and UMajor, the
three lowest luminosity Galactic dSphs, all have the same dispersion,
about $\vdisp=10\,\kms$, and And IX, the second faintest known dSph,
has $\vdisp=6.8^{+3.0}_{-2.0}$ \citep{Chapman05}.

Is there a minimum halo size in which stars can form, or a minimum
clustering scale for the dark matter?  There was no survivability
reason why UMajor could not have been much less massive.  For
instance, the mass required to bind a $500\,\rm pc$ dwarf halo against
tidal disruption by a $10^{12}\,\Msun$ Galaxy at UMajor's distance of
$100\,\rm\kpc$ is estimated by  equating the dwarf and Galactic
densities, so that $M_{\rm bind}\sim 10^{12}\,\Msun\times (500\,{\rm
pc}/100\,{\rm kpc})^3 \sim 10^5\,\Msun$, or about two orders of
magnitude less massive than UMajor's central region, and perhaps three
orders of magnitude less massive than its global mass.

\citet{Kormendy04} have compiled  scaling relations
among dwarf galaxies and more massive ellipticals, relating $M_B$,
$\rho_0$, $r_{\rm core}$, and $\vdisp$. We can locate UMajor in this
ensemble, using the values $\rho_0=0.18\,\Msun \,\rm pc^{-3}$,
$\vdisp=9.3\,\kms$, $r_{\rm core}\sim r_{1/2}\sim 250\,\rm pc$, and
$M_B=M_V+0.9=-5.65$, where $M_V=-6.75$ has been converted to $B$ band
using the typical $B-V\approx0.9$ color of a K-giant.  From Figures 2
and 4 of Kormendy \& Freeman (2004), we find, unsurprisingly, that
UMajor is typical of the dSphs in all respects but luminosity.  All
but two of the previously known dSphs and dIrrs fall onto a tight
Faber-Jackson relation between velocity dispersion and luminosity,
albeit with a systematic offset relative to larger ellipticals.  UMi
and Draco are the exceptions, being about three magnitudes
under-luminous for their mass.  To agree with the general trend,
UMajor would need to have a dispersion of $<3\,\kms$,
or would have to be about five magnitudes brighter. 


It has been suggested that the dSphs' extreme $M/L$ results from the
expulsion of baryons from their shallow potential well by
supernovae \citep{Larson74,Dekel86}, Because the potential wells of
the dSphs seem comparable, UMajor must have been intrinsically baryon
poor to begin with, or else the baryon expulsion efficiency must have
varied by one or two orders of magnitude among dSphs in order to
produce their very different luminosities.  
UMajor demonstrates that the luminosity scatter at the low mass end of the
galaxy distribution is very large. It appears likely that more dark
and massive dwarfs are lurking in the vicinity of the Galaxy. Detections
of more candidate objects are urgently needed to check whether the number
and properties of the population are truly consistent with the missing
dark matter haloes of the simulations.

\vskip 2cm

\acknowledgments

The data presented herein were obtained at the W.M. Keck Observatory,
which is operated as a scientific partnership among the California
Institute of Technology, the University of California and the National
Aeronautics and Space Administration. The Observatory was made
possible by the generous financial support of the W.M. Keck
Foundation.

JTK gratefully acknowledges the support of the Beatrice Watson Parrent
Fellowship.  MIW acknowledges the support of PPARC.

\eject

\begin{deluxetable}{cllrrrrrr}
\tablecolumns{8}
\tablecaption{\label{table:stars}Positions, magnitudes, and velocities of UMajor candidate stars}
\tablehead{
\colhead{Object} & \colhead{RA} & \colhead{Dec} & 
  \colhead{$g$} &   \colhead{$i$} &
  \colhead {$v$} &  \colhead {$\sigma_{\rm IRAF}$} &
  \colhead {$\sigma_{\rm scat}$} \\
\colhead{} & \multicolumn{2}{c}{(J2000)} &
  \colhead{} &\colhead{} &
   \colhead {($\kms$)} &  \colhead {($\kms$)} &
  \colhead {($\kms$)} 
 }
\startdata
1 & $10^{\rm h}35^{\rm m}28$\rlap{$^{\rm s}$}.$51$ & $+51^{\circ}57^{\prime}00$\rlap{$^{\prime\prime}$}.$9$  & 18.64 & 17.45 & -54.87 & 1.00 & 1.30 \\
2 & $10^{\rm h}34^{\rm m}30$\rlap{$^{\rm s}$}.$51$ & $+51^{\circ}57^{\prime}07$\rlap{$^{\prime\prime}$}.$0$  & 18.88 & 17.75 & -50.80 & 0.96 & 1.09 \\
3 & $10^{\rm h}35^{\rm m}15$\rlap{$^{\rm s}$}.$87$ & $+51^{\circ}59^{\prime}32$\rlap{$^{\prime\prime}$}.$0$  & 18.93 & 17.82 & 15.12 & 1.45 & 0.94 \\
4 & $10^{\rm h}34^{\rm m}52$\rlap{$^{\rm s}$}.$44$ & $+51^{\circ}57^{\prime}02$\rlap{$^{\prime\prime}$}.$2$  & 19.08 & 17.89 & 8.21 & 1.50 & 0.40 \\
5 & $10^{\rm h}34^{\rm m}52$\rlap{$^{\rm s}$}.$05$ & $+51^{\circ}58^{\prime}28$\rlap{$^{\prime\prime}$}.$3$  & 19.29 & 18.25 & -64.22 & 1.90 & 0.65 \\
6 & $10^{\rm h}34^{\rm m}42$\rlap{$^{\rm s}$}.$36$ & $+51^{\circ}58^{\prime}06$\rlap{$^{\prime\prime}$}.$1$  & 19.45 & 18.37 & -38.95 & 1.80 & 1.25 \\
7 & $10^{\rm h}35^{\rm m}17$\rlap{$^{\rm s}$}.$23$ & $+51^{\circ}55^{\prime}33$\rlap{$^{\prime\prime}$}.$7$  & 19.48 & 18.38 & -53.53 & 5.00 & 5.15 \\
\enddata
\tablecomments{ First column is an object identification number. Columns 2 and 3 are the position in J2000 coordinates. 
Columns 4 and 5 are the SDSS $g$ and $i$ magnitudes.  Columns 6, 7,
and 8 are $v$, the measured heliocentric velocity; $\sigma_{\rm
IRAF}$, the uncertainty obtained from the rescaled {\sc IRAF} {\sc
fxcor} velocity error; and $\sigma_{\rm scat}$, the uncertainty
obtained from the velocity scatter among the different absorption
lines.}
\end{deluxetable}


\begin{thebibliography}{}
%
\bibitem[Abazajian et al.(2005)]{ABA05} 
Abazajian, K., Adelman, J., Agueros, M., et al. 2005, AJ, 129, 1755
%
\bibitem[Armandroff \& Da Costa(1991)]{ARM91} Armandroff, T.~E., 
 \& Da Costa, G.~S., 1991, AJ, 101, 1329
%
\bibitem[Bahcall \& Tremaine(1981)]{Bahcall81}Bahcall, J., \& Tremaine, S., 1981, ApJ, 244, 805
%
\bibitem[Blitz et al.(1999)]{Blitz99}Blitz, L., Spergel, D.~N., 
 Teuben, P.J., Hartmann, D., \& Burton, W.B., 1999, ApJ, 514, 818
%
\bibitem[Chapman et al.(2005)]{Chapman05}Chapman, S.~C., Ibata, R., Lewis, G.~R., 
 Ferguson, A.~M.~N., Irwin, M., McConnachie, A., \& Tanvir, N.,
 2005, accepted for publication in ApJL (astro-ph/0506103)
%
\bibitem[Dekel \& Silk(1986)]{Dekel86} Dekel,A., \& Silk, J., 1986, ApJ, 303, 39
%
\bibitem[Evans et al.(2003)]{Evans03} Evans, N.~W., Wilkinson, M.~I., Perrett, K.~M., \& Bridges, T.~J.\ 2003, \apj, 583, 752 
%
%
\bibitem[Heisler, Tremaine, \& Bahcall(1985)]{Heisler85} Heisler, J., Tremaine, S., \& Bahcall, J.~N.\ 1985, \apj, 298, 8
%
\bibitem[de Heij, Braun, \& Burton (2002)]{deHeij02}
de Heij V., Braun R., Burton W.B., 2002, AA, 392, 417 
%
\bibitem[Kleyna et al.(2001)]{KleynaDraco} Kleyna, J.~T., Wilkinson, M.~I., Evans, N.~W., \& Gilmore, G.\ 2001, ApJL, 563, L115 
%
\bibitem[Kleyna et al.(2004)]{KleynaSextans}Kleyna J.~T., Wilkinson M.~I., 
         Evans N.W., Gilmore G., 2004, MNRAS, 354, L66
%
\bibitem[Klypin et al.(1999)]{Klypin99} Klypin, A., Kravtsov, A.~V., Valenzuela, O., \& Prada, F.\ 1999, ApJ, 522, 82 
%
\bibitem[Kormendy \& Freeman(2004)]{Kormendy04} Kormendy, J., 
 \& Freeman, K.~C.\ 2004, IAU Symposium, 220, 377 
%
\bibitem[Larson(1974)]{Larson74}Larson, R.B., 1974, MNRAS, 169, 229
%
\bibitem[Mateo(1998)]{MAT98} Mateo M., 1998, ARA\&A, 36, 435
%
\bibitem[Mateo et al.(1998)]{MateoLeo98} Mateo, M., Olszewski, E.~W., Vogt, S.~S., \& Keane, M.~J.\ 1998, \aj, 116, 2315 
%
\bibitem[Meurer et al.(1996)]{Meurer96} Meurer, G.~R., Carignan, C., Beaulieu, S.~F., \& Freeman, K.~C.\ 1996, \aj, 111, 1551
%
\bibitem[Minchin et al.(2005)]{Minchin05} Minchin, R., et al.\ 
2005, \apjl, 622, L21 
%
\bibitem[Navarro, Frenk, \& White(1997)]{NFW97} Navarro, J.~F., Frenk, C.~S., \& White, S.~D.~M.\ 1997, ApJ, 490, 493 
%
\bibitem[Richstone \& Tremaine(1986)]{RT86}Richstone, D.O., \& Tremaine, S., 
1986, AJ, 92, 72
%
 \bibitem[Thilker et al.(2004)]{Thilker04} Thilker, D.~A., Braun, R., 
Walterbos, R.~A.~M., Corbelli, E., Lockman, F.~J., Murphy, E., \& 
Maddalena, R.\ 2004, 
\apjl, 601, L39 
%
\bibitem[Wilkinson et al.(2004)]{WilkDracoUMi} Wilkinson, M.~I., Kleyna, 
J.~T., Evans, N.~W., Gilmore, G.~F., Irwin, M.~J., \& Grebel, E.~K.\ 2004, ApJ, 611, L21 
%
\bibitem[Willman et al.(2005)]{Willman05}  Willman, B., et al., 2005, ApJL, submitted, (astro-ph/0503552)
%
\bibitem[Vogt (1994)]{Vogt94} Vogt, S. S., et al. 1994, Proc. SPIE, 2198, 362
%
\bibitem[Zucker et al.(2004)]{Zucker04} Zucker, D.~B., et al.\ 
2004, \apjl, 612, L121 
 
%
\end{thebibliography}
\end{document}